\documentclass{ws-ijqi-mod}

\usepackage{graphicx}
\usepackage{amssymb,bbm}

\newcommand{\ket}[1]{|#1\rangle}
\newcommand{\bra}[1]{\langle#1|}
\newcommand{\braket}[1]{\langle#1\rangle}
\newcommand{\rme}{\mathrm{e}}
\newcommand{\ud}{\mathrm{d}}

\newcommand{\Id}{\mathbbm{1}}
\newcommand{\complex}{\mathbbm{C}}
\newcommand{\pFq}[1]{\!\,_{#1}F}

\begin{document}

\markboth{A. G\'abris and G. S. Agarwal}
{Quantum teleportation with pair-coherent states}

%%%%%%%%%%%%%%%%%%%%% Publisher's Area please ignore %%%%%%%%%%%%%%
\catchline{}{}{}{}{}
%%%%%%%%%%%%%%%%%%%%%%%%%%%%%%%%%%%%%%%%%%%%%%%%%%%%%%%%%%%%%%%%%%%

\title{QUANTUM TELEPORTATION WITH PAIR-COHERENT STATES}

\author{AUR\'EL G\'ABRIS}

\address{Department of Nonlinear and Quantum Optics, \\
Research Institute for Solid State Physics and Optics, 
Hungarian Academy of Sciences,\\
Budapest P.O. Box 49, 1525 Hungary\\
gabrisa@szfki.hu}

\author{GIRISH S. AGARWAL%
\footnote{On leave of absence from Physical Research Laboratory,
  Navrangpura, Ahmedabad 380 009, India.}}

\address{
Department of Physics, Oklahoma State University,\\
Stillwater, OK 74078, USA\\
girish.agarwal@okstate.edu}

\maketitle

\begin{history}
%\received{Day Month Year}
%\revised{Day Month Year}
%\accepted{Day Month Year}
%\comby{(xxxxxxxxxx)}
\end{history}

\begin{abstract}
  Recently it has been argued that all presently performed continuous
  variable quantum teleportation experiments could be explained using
  a local hidden variable theory. In this paper we study a
  modification of the original protocol which requires a fully quantum
  mechanical explanation even when coherent states are teleported. Our
  calculations of the fidelity of teleportation using a pair-coherent
  state under ideal conditions suggests that fidelity above the
  required limit of $1/2$ may be achievable in an experiment also.
\end{abstract}


%%%%%%%%%%%%%%%%%%%%%%%%%%%%%%%%%%%%%%%%%%%%%%%%%%%%%%%%%%%%%%%%%%%%%
\section{Introduction}

In their recent paper Caves \textit{et al.}\cite{Caves-prl93} argued
that the process of continuous variable
teleportation\cite{Braunstein-prl80} of Gaussian states allows for a
local hidden variable description. Earlier
works\cite{Braunstein-jmo47,Braunstein-pra64} focused on the
necessity of entanglement for teleportation, and found that
fidelities up to $\mathcal{F}_{\mathrm{av}}=1/2$ can be achieved even with no
entanglement.

In the present paper we focus on a teleportation scheme that does not
allow for a local hidden variable description. In
Ref.~\refcite{Caves-prl93} it has been suggested to use non-Gaussian
states at the input. For this case, an upper limit
$\mathcal{F}_{\mathrm{av}}=2/3$  on the fidelity has been
established, below which there may exist a local hidden variable
description.

Experimental realization of such a scheme involving bright
non-Gaussian input states, however, does not seem formidable in the
near future. The underlying problem is that to achieve such high
fidelity, we would require high entanglement, and therefore high
photon numbers for both the entangled resource and the input.

On the other hand, replacing the entangled resource with a
non-Gaussian state can also render the local realistic description
impossible. In the present paper we propose to use a
pair-coherent\cite{Agarwal-josab5,Agarwal-prl57} state as an entangled
resource. In contrast to the two-mode squeezed vacuum for which
measurement of commuting quadrature operators do not violate local
realism,\cite{Banaszek-pra58,Banaszek-prl82,ZBChen-prl88} such
measurements on a pair-coherent state cannot be described by a local
hidden variable theory.\cite{Munro-pra59} And since the teleportation
protocol calls for measurement of commuting quadratures, this
replacement is necessary to rule out a local realistic description of
teleportation of Gaussian states.

The use of pair-coherent states seems viable due to recent
developments in experiments with non-degenerate optical parametric
oscillators\cite{Hayasaka-ol29,JGao-ol23} (NOPO), which have been
shown theoretically to produce pair-coherent states under certain
conditions.\cite{Reid-pra47,Gilchrist-joptb2} Since the pair-coherent
state is non-Gaussian, the negative part of its Wigner function has
non-zero support, and this is enough to eliminate the possibility of a
local hidden variable description of the teleportation process.

The structure of our paper is as follows: In
Sec.~\ref{sec:pair-coherent} we briefly recall the definition and
important properties of pair-coherent states. In
Sec.~\ref{sec:teleport-coherent} we overview our formalism used to
describe the teleportation process and present the results on the
teleportation fidelities of coherent states.  In
Sec.~\ref{sec:discussion} we interpret our results in the light of the
earlier requirements, and we finally conclude in
Sec.~\ref{sec:conclusions}.

%%%%%%%%%%%%%%%%%%%%%%%%%%%%%%%%%%%%%%%%%%%%%%%%%%%%%%%%%%%%%%%%%%%%%
\section{Pair-coherent states}
\label{sec:pair-coherent}

Pair-coherent states\cite{Agarwal-josab5} of the two mode
electromagnetic field are simultaneous eigenstates of the
pair-annihilation operator $ab$ and the difference of the number
operators $a^{\dag}a - b^{\dag}b$, i.~e.
\begin{equation}
ab \ket{\zeta,q} = \zeta\ket{\zeta,q}, \qquad\mbox{and}\qquad
a^{\dag}a - b^{\dag}b \ket{\zeta,q} = q\ket{\zeta,q}.
\end{equation}
These states are important in quantum information processing, because
they are the only non-Gaussian entangled states\cite{Agarwal-joptb7}
that could be prepared reliably sometime in the near
future.\cite{Reid-pra47,Gilchrist-joptb2,Hayasaka-ol29,JGao-ol23}

Pair-coherent states may either be written using the Fock basis as
\begin{equation}
\ket{\zeta,q} = \frac{1}{\sqrt{\zeta^q I_0(2\zeta)} } \sum
\frac{\zeta^n}{\sqrt{n!(n+q)!}} \ket{n+q,n},
\label{eq:pair-coherent-num-q}
\end{equation}
or in a coherent state representation as
\begin{equation}
\ket{\zeta,q} =   \frac{\rme^{\zeta}}{2\pi\sqrt{\zeta^q I_0(2\zeta)} }
  \int_{0}^{2\pi} \left(\sqrt{\zeta}\rme^{i\vartheta}\right)^{-q}
\ket{\sqrt{\zeta}\rme^{i\vartheta}} \ket{\sqrt{\zeta}\rme^{-i\vartheta}}
\ud\vartheta,
\label{eq:pair-coherent-q}
\end{equation}
where $I_0$ denotes the modified Bessel function of the first kind.
Since Eq.~(\ref{eq:pair-coherent-q}) is essentially a contour integral
on a zero centered circle of the complex plane, pair-coherent states
are also sometimes referred to as ``circle'' states.

In the present paper we make use of two features of pair-coherent
states.  First is that they are entangled whenever $\zeta>0$, as can
easily be seen from Eq.~(\ref{eq:pair-coherent-num-q}), which happens
to be a valid Schmidt decomposition. The other is that these states
are also non-Gaussian with their Wigner functions exhibiting
significant negativity:
\begin{multline}
W(\alpha/2; \beta/2) = \frac{4 \exp(-|\alpha|^2 - |\beta|^2)}{ \pi^2
  I_0(2\zeta)} \sum_{m,n=0}^{\infty} \cos[ (m-n) (\varphi_{\alpha} +
  \varphi_{\beta})] \frac{ (\zeta |\alpha| |\beta|)^{m+n}}{ (m!)^2
  (n!)^2} \\
\times \pFq{2}_0(-m,-n;;-1/|\alpha|^2) \pFq{2}_0(-m,-n;;-1/|\beta|^2),
\label{eq:pair-coherent-wig}
\end{multline}
where $\varphi_{\alpha}$ and $\varphi_{\beta}$ denote the argument of
$\alpha$ and $\beta$, respectively. (The generalized hypergeometric
function $\pFq{2}_0$ represents only a finite sum.)

%%%%%%%%%%%%%%%%%%%%%%%%%%%%%%%%%%%%%%%%%%%%%%%%%%%%%%%%%%%%%%%%%%%%%
\section{Teleportation of coherent states}
\label{sec:teleport-coherent}

Now we turn to the modified quantum teleportation protocol, where the
two-mode squeezed vacuum constituting the quantum channel has been
replaced by a pair-coherent state $\ket{\zeta,0}$.  Comparing
Eq.~(\ref{eq:pair-coherent-wig}) with
Eqs.~(\ref{eq:pair-coherent-num-q}) and (\ref{eq:pair-coherent-q}), we
find that the description of pair-coherent states is more convenient
in the Hilbert space formalism. Indeed, when the input states are
coherent states, the coherent state representation
(\ref{eq:pair-coherent-q}) turns out to be the most efficient.

We use the coherent state description of Bell state measurement from
Ref.~\refcite{Janszky-pra64}. We write the state corresponding to
Alice's homodyne measurement outcome $A=(X+iP)/\sqrt2$ as
\begin{equation}
\ket{B(X,P)} = \int_{\complex} \rme^{A\gamma^* - A^*\gamma}
\ket{\gamma+A} \ket{\gamma^*-A^*}  \,\ud^2\gamma.
\end{equation}
The classical information sent to Bob corresponds to this complex
value $A$.

We construct the transfer operator following
Ref.~\refcite{Hofmann-pra62} using this expression. To obtain the
transfer operator in general terms, we first note that the Bell states
possess the property
\begin{equation}
\ket{B(X,P)} = D(A) \otimes D(-A^*) \ket{B(0,0)} = 
D(2A) \otimes \Id \ket{B(0,0)},
\end{equation}
which together with $_1\!\braket{\psi|B(0,0)}_{12} = \mathcal{I}
\ket{\psi}_2$ imply
\begin{equation}
_2\!\braket{\psi|B(X,P)}_{12} = \mathcal{I} D(2A)\ket{\psi}_1,
\end{equation} 
where $\mathcal{I}$ is the anti-linear operator primitive defined as
$\mathcal{I}\ket{n} = \ket{n}$. Since this partial scalar product
describes Alice's measurement, the anti-linearity enters into the
expression of the transfer operator:
\begin{equation}
T_{\zeta}(A) := \frac{\rme^{\zeta}}{2\pi\sqrt{\pi I_0(2\zeta)}} \int_0^{2\pi}
\ket{\sqrt{\zeta}\rme^{i\vartheta}} \bra{\sqrt{\zeta}\rme^{i\vartheta}} 
D(-2A) \,\ud\vartheta.
\end{equation}

The teleportation procedure can be written as follows. Our initial
state is $\ket{\psi_0}_{123} = \ket{\psi_{\mathrm{in}}}_1
\ket{\zeta,0}_{23}$, and the transfer operator gives the unnormalized
outcome
\begin{equation}
\ket{\psi_t(A)}=T_{\zeta}(A)\ket{\psi_{\mathrm{in}}}
\label{eq:psi1-prop}
\end{equation}
of the first stage of the teleportation --- before Bob's displacement. We
note that this definition is slightly different from that of
Ref.~\refcite{Hofmann-pra62}. The probability distribution for Alice's
measurement is given by the norm $P(A)=\|\ket{\psi_t(A)}\|$. Hence the
actual output state after the displacement can be written
$
\ket{\psi_{\mathrm{out}}(A)} = 1/\sqrt{P(A)} D(\beta) T_{\zeta}(A)
\ket{\psi_{\mathrm{in}}(A)}
$
if we assume that Bob attempts to reconstruct the state with a
coherent displacement $\beta$. We use the definition of average fidelity
\begin{equation}
\mathcal{F}_{\mathrm{av}} = \int_{\complex} \mathcal{F}(A) P(A)
\,\ud^2A
\label{eq:avg-fid}
\end{equation}
where $\mathcal{F}(A) = |\braket{\psi_{\mathrm{in}}|D(\beta)
  T_{\zeta}(A) | \psi_{\mathrm{in}}}|^2/P(A)$ is the fidelity of a
single teleportation event. In the following, we shall take $\beta =
g2A$, and assume that the gain factor $g$ may be adjusted by Bob to
improve the fidelity of teleportation.

\begin{figure}[pb]
\begin{center}
\includegraphics[scale=0.71]{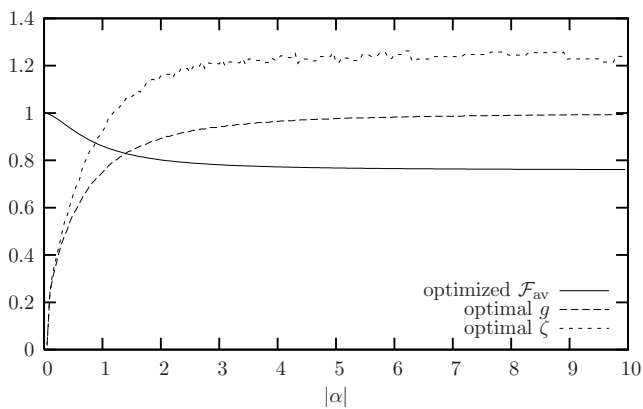}
\caption{Optimized average fidelity plotted against the input state
  $\alpha$. Also plotted are the optimal values of $\zeta$ and $g$.
  The plots demonstrate that the optimal gain $g\to1$ as
  $|\alpha|\to\infty$, hence $\zeta\to1.2357$ approximately in the same
  limit. \label{fig:avgfid-coh-max}}
\end{center}
\end{figure}

\begin{figure}[pb]
\begin{center}
\includegraphics[scale=0.71]{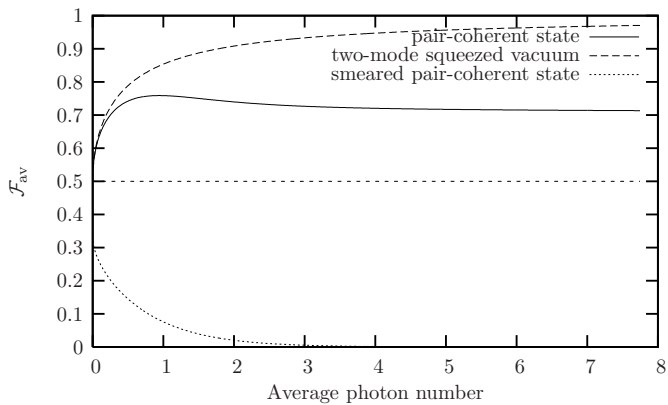}
\caption{Average fidelities for the teleportation of coherent states
  using a two-mode squeezed state and a pair coherent state, when the
  gain tuning $g=1$. In both cases the average fidelity is independent
  of the input coherent state $\ket{\alpha}$. Also plotted is the
  fidelity of teleportation when using a ``smeared-out'' pair-coherent
  state. \label{fig:avgfid-coh-g1}}
\end{center}
\end{figure}

Now let the input state be $\ket{\psi_{\mathrm{in}}} = \ket{\alpha}$,
a coherent state.  According to Eq.~(\ref{eq:psi1-prop}), after
Alice's measurement our pseudo output state is
\begin{equation}
\ket{\psi_t(A)} =  \frac{\rme^{\zeta}}{2\pi \sqrt{\pi I_0(2\zeta)}} \int
\braket{\sqrt{\zeta} \rme^{i\vartheta}| D(-2A) 
  |\alpha} \ket{\sqrt{\zeta} \rme^{i\vartheta}},
\end{equation}
which gives for the average fidelity the expression
\begin{multline}
\mathcal{F}_{\mathrm{av}}(\alpha,\zeta,g) =
  \frac{\rme^{-\frac{(g-1)^2}{1+g^2}|\alpha|^2}} {4 
  (1+g^2) I_0(2\zeta)} 
\sum_{m,n=0}^{\infty} \frac{\zeta^{m+n}}{(m!n!)^2}
\sum_{p=0}^{m} \sum_{q=0}^{n}
{m \choose p}{n \choose q}  (-1)^{m+n+p+q} \\
\times
\sum_{j=0}^{\min\{n+p,m+q\}} {n+p \choose j}{m+q \choose j} j!
\left(\frac{(1-g)^2}{1+g^2}\right)^{m+n-j}
  \frac{g^{m+n+2(p+q-j)}}{(1+g^2)^{p+q}}  (|\alpha|^2)^{m+n-j}.
\end{multline}
In the special case $g=1$ this simplifies to
\begin{equation}
\mathcal{F}_{\mathrm{av}}(\alpha,\zeta,1) = \frac1{2I_0(2\zeta)}
\sum_{n,m=0}^{\infty}  \frac{(\zeta/2)^{m+n} (m+n)!}{(m!n!)^2},
\end{equation}
and hence becomes independent of the input coherent amplitude
$\alpha$.  This fidelity is compared with that of the original
protocol on Fig.~\ref{fig:avgfid-coh-g1}. On
Fig.~\ref{fig:avgfid-coh-max} we can observe that the effect of gain
tuning is similar to that of on the original protocol: the optimal
gain is always $g_{\mathrm{opt}}<1$, and it approaches $1$ in the
limit of large amplitudes.

%%%%%%%%%%%%%%%%%%%%%%%%%%%%%%%%%%%%%%%%%%%%%%%%%%%%%%%%%%%%%%%%%%%%%
\section{Discussion}
\label{sec:discussion}

An interesting feature of the pair-coherent state quantum channel is
that the fidelity does not always increase with the entanglement,
which is reflected in a maximum of the fidelity as a function of
$\zeta$. The likely reason for this is that the Bell measurement
consisting of the beam-splitter and the homodyne detectors is not the
matching\cite{Kurucz-joptb5} measurement for the pair-coherent state.
This is in great contrast with the two-mode squeezed vacuum, for which
this measurement becomes matching in the limit of infinite squeezing.
As indicated by the numerical results, the parameters are probably the
closest to the matching conditions when the pair-coherent state has
$\zeta=1.2357$.

As $\zeta$ is increased, the fidelity appears to approach a limit
which is above $1/2$. Following Ref.~\refcite{Braunstein-pra64} we
conclude that the entanglement contained in the pair-coherent state is
being utilized in the process. In addition, we argue that obtaining
fidelity greater than $1/2$ also rules out the existence of an
extended hidden variable model suggested by Caves and
W\'odkiewicz.\cite{Caves-prl93} This can be verified by calculating
the teleportation fidelity using the ``smeared-out'' pair-coherent
state $W_{\mathrm{kicked}}(\alpha,\beta) = Q(\alpha,\beta)
\propto \exp(-|\alpha|^2-|\beta|^2)
|I_0(2\sqrt{\zeta\alpha,\beta})|^2$. This fidelity is also included on
Fig.~\ref{fig:avgfid-coh-g1}, and shows that Alice and Bob obtain very
low fidelities even for $\zeta=0$, i.~e.\ when they are sharing the
vacuum. However, they have no other choice than applying the maximum
``kicking'' strength $t=1$, since the case $\zeta=0$ is of zero
measure, and for all other $\zeta$ they must resort to the $Q$
function of the original.\cite{Lutkenhaus-pra51}

%%%%%%%%%%%%%%%%%%%%%%%%%%%%%%%%%%%%%%%%%%%%%%%%%%%%%%%%%%%%%%%%%%%%%
\section{Conclusions}
\label{sec:conclusions}

In this paper we have proposed a modified setup of continuous variable
teleportation which may allow for an experimental test of local
realism in the teleportation process. Instead of the approach of
Ref.~\refcite{Caves-prl93}, we replaced the quantum channel with a
non-Gaussian, pair-coherent state. This permits us to use ordinary
coherent states on the input and still rule out the possibility of a
local hidden variable description. Teleporting coherent states has the
additional advantage that the experimental determination of fidelity
may be easier compared to some non-Gaussian input.

Our calculations yield a maximum fidelity
$\mathcal{F}_{\mathrm{av}}=0.75884$ for a particular pair-coherent
state. This seems to be distant enough from $1/2$, therefore
fidelities exceeding the classical limit could be observed
experimentally also.  We have also shown that this setup rules out the
existence of an extended hidden-variable model\cite{Caves-prl93} based
on smearing out the participating quantum states.

%%%%%%%%%%%%%%%%%%%%%%%%%%%%%%%%%%%%%%%%%%%%%%%%%%%%%%%%%%%%%%%%%%%%%
\section*{Acknowledgments}
This work was partly supported by the National Research Fund of
Hungary under contract Nos.\ T~043079 and T~049234. A.~G.\ 
would also like to thank J.\ Janszky for useful discussions.

%\bibliographystyle{ws-van}
%\bibliography{pair-coherent}

\begin{thebibliography}{19}
\providecommand{\natexlab}[1]{#1}
\providecommand{\url}[1]{\texttt{#1}}
\expandafter\ifx\csname urlstyle\endcsname\relax
  \providecommand{\doi}[1]{doi: #1}\else
  \providecommand{\doi}{doi: \begingroup \urlstyle{rm}\Url}\fi

\bibitem{Caves-prl93}
C.~M. Caves and K.~W\'odkiewicz, \emph{Phys. Rev. Lett.} {\bf 93} (2004),
  \penalty0 040506.

\bibitem{Braunstein-prl80}
S.~L. Braunstein and H.~J. Kimble, \emph{Phys. Rev. Lett.} {\bf 80} (1998),
  \penalty0 869--872.; L. Vaidman, \emph{Phys. Rev. A} {\bf 49} (1994)
  \penalty0 1473--1476.

\bibitem{Braunstein-jmo47}
S.~L. Braunstein, C.~A. Fuchs, and H.~J. Kimble, \emph{J. Mod. Opt}. {\bf 47}
  (2000), \penalty0 267--278.

\bibitem{Braunstein-pra64}
S.~L. Braunstein, C.~A. Fuchs, H.~J. Kimble, and P.~van Loock, \emph{Phys. Rev.
  A}. {\bf 64} (2001), \penalty0 022321.

\bibitem{Agarwal-josab5}
G.~S. Agarwal, \emph{J. Opt. Soc. Am. B}. {\bf 5} (1988), \penalty0 1940--1947.

\bibitem{Agarwal-prl57}
G.~S. Agarwal, \emph{Phys. Rev. Lett.} {\bf 57} (1986), \penalty0 827--830.

\bibitem{Banaszek-pra58}
K.~Banaszek and K.~W\'odkiewicz, \emph{Phys. Rev. A}. {\bf 58} (1998),
  \penalty0 4345--4347.

\bibitem{Banaszek-prl82}
K.~Banaszek and K.~W\'odkiewicz, \emph{Phys. Rev. Lett.} {\bf 82} (1999),
  \penalty0 2009--2013.

\bibitem{ZBChen-prl88}
Z.-B. Chen, J.-W. Pan, G.~Hou, and Y.-D. Zhang, \emph{Phys. Rev. Lett.} {\bf
  88} (2002), \penalty0 040406.

\bibitem{Munro-pra59}
W.~J. Munro, \emph{Phys. Rev. A}. {\bf 59} (1999), \penalty0 4197--4201.

\bibitem{Hayasaka-ol29}
K.~Hayasaka, Y.~Zhang, and K.~Kasai, \emph{Opt. Lett.} {\bf 29} (2004),
  \penalty0 1665--1667.

\bibitem{JGao-ol23}
J.~Gao, F.~Cui, C.~Xue, C.~Xie, and P.~Kunchi, \emph{Opt. Lett.} {\bf 23}
  (1998), \penalty0 870--872.

\bibitem{Reid-pra47}
M.~D. Reid and L.~Krippner, \emph{Phys. Rev. A}. {\bf 47} (1993), \penalty0
  552--555.

\bibitem{Gilchrist-joptb2}
A.~Gilchrist and W.~J. Munro, \emph{J. Opt. B}. {\bf 2} (2000), \penalty0
  47--52.

\bibitem{Agarwal-joptb7}
G.~S. Agarwal and A.~Biswas, \emph{J. Opt. B: Quantum Semiclass. Opt.} {\bf 7}
  (2005), \penalty0 350--354.

\bibitem{Janszky-pra64}
J.~Janszky, M.~Koniorczyk, and A.~G\'abris, \emph{Phys. Rev. A}. {\bf 64}
  (2001), \penalty0 034302.

\bibitem{Hofmann-pra62}
H.~F. Hofmann, T.~Ide, and T.~Kobayashi, \emph{Phys. Rev. A}. {\bf 62} (2000),
  \penalty0 062304.

\bibitem{Kurucz-joptb5}
Z.~Kurucz, M.~Koniorczyk, P.~Adam, and J.~Janszky, \emph{J. Opt. B: Quantum
  Semiclass. Opt.} {\bf 5} (2003), \penalty0 S627--S632.

\bibitem{Lutkenhaus-pra51}
N.~L\"utkenhaus and S.~M. Barnett, \emph{Phys. Rev. A}. {\bf 51} (1995),
  \penalty0 3340--3342.

\end{thebibliography}

\end{document}